\documentclass[12pt,epsfig]{article}
\newcommand{\cur}{\mbox{\scriptsize cur}}

\newcommand{\eqr}[1]{Eq.~(\ref{#1})}
\newcommand{\eqrs}[2]{Eqs.~(\ref{#1}) and (\ref{#2})}
\begin{document}
\title{\bf
Small Strange Quark Content of Protons}
\author{
Michael D.\ Scadron\footnote{Electronic address: scadron@physics.arizona.edu}\\
{\normalsize\it Physics Department, University of Arizona, Tucson, AZ 85721,
USA} \\[3mm]
\and
Frieder Kleefeld\footnote{Collaborator of \em
Centro de F\'{\i}sica das Interac\c{c}\~{o}es Fundamentais; \em
electronic address: kleefeld@cfif.ist.utl.pt} \\
{\normalsize\it Doppler Institute for Mathematical Physics and Applied
Mathematics \& Nuclear Physics Institute} \\ {\normalsize\it Czech Academy of
Sciences, 250\,68 \v{R}e\v{z}, Czech Republic} \\[3mm]
\and
George Rupp\footnote{Electronic address: george@ist.utl.pt}  \\
{\normalsize\it Centro de F\'{\i}sica das Interac\c{c}\~{o}es Fundamentais,
Instituto Superior T\'{e}cnico}\\ {\normalsize\it P-1049-001 Lisboa, Portugal}
%\and
%E.\ van Beveren\footnote{Electronic address: eef@teor.fis.uc.pt} \\
%{\normalsize\it Centro de F\'{\i}sica Te\'{o}rica, Departamento de F\'{\i}sica,
%Universidade de Coimbra} \\ {\normalsize\it P-3004-516 Coimbra, Portugal}\\[3mm]
}
\date{\today}
\maketitle

\begin{abstract}
The contribution of strange sea quarks to the proton mass and spin, as well
as the related pion-nucleon sigma term, are briefly revisited, in the light
of new experimental and lattice results. Also the predictions of chiral
perturbation theory for these quantities are discussed.  \\[5mm]
\end{abstract}

Very recent combined experimental \cite{T06a} and lattice \cite{T06b} results
now unmistakably show that the strange-quark sea contribution to the proton
mass is very small, suggesting \cite{T06c}
\begin{equation}
\Delta m_N^{(s)} \; = \; \frac{y}{2}\left(\frac{m_s}{\hat{m}}\right)_{\!\!\cur}
\sigma_{\pi N} \; \sim \; 10\;\mbox{MeV}\footnote{Somewhat higher values, up
to 30 MeV, cannot be completely excluded \cite{T06d}} \; ,
\label{dmns}
\end{equation}
where $\hat{m}=(m_u+m_d)/2$, and the strangeness quark ratio $y$ is defined as
\begin{equation}
y\;\equiv\;\frac{2\langle p|\bar{s}s|p\rangle}{\langle p|\bar{u}u+
\bar{d}d|p \rangle} \; . 
\label{ydef}
\end{equation}
On the light plane, it was first estimated that \cite{SS74}
\begin{equation}
\left(\frac{m_s}{\hat{m}}\right)_{\!\cur} \; \approx \; 6\,\mbox{--}\,7 \; .
\label{lightplane}
\end{equation}
Using \eqrs{ydef}{lightplane} together with the usual pion-nucleon
$\sigma$-term $\sigma_{\pi N}\approx65$ MeV, \eqr{dmns} can then be solved for
$y$ as
\begin{equation}
y \; = \; \frac{2\Delta m_N^{(s)}}{\displaystyle\left(\frac{m_s}
{\hat{m}}\right)_{\!\!\cur}\sigma_{\pi N}} \;\sim\;
 \frac{2\times10\:\mbox{MeV}}{6.5\times65\:\mbox{MeV}} \; \sim \; 5\% \; .
\label{ysmall}
\end{equation}

In fact, 13 years ago the measurement of the nucleon strange-quark
distribution, using neutrino charm production \cite{R93}, found a comparable
result, viz.\
\begin{equation}
y \; = \; 0.064^{+0.008}_{-0.007} \; .
\label{yexp}
\end{equation}
Moreover, the Feynman-Hellman theorem \cite{F39} led to nearly
model-independent results of 7--10\% \cite{K88} and $\leq5$\% \cite{SC90}.
Recently, the pion and kaon $\bar{q}q$ masses were dynamically generated,
again leading to a similar ratio \cite{SKR06}
\begin{equation}
y \; \approx \; 5\,\mbox{--}\,6\,\mbox{\%} \; .
\label{yskr}
\end{equation}
Lastly, scalar $\sigma$ and $f_0(980)$ tadpole leakage gives rise to about
the same scale \cite{S92}, i.e.,
\begin{equation}
y \; \sim \; 6\,\mbox{\%} \; ,
\label{ytadpole}
\end{equation}
with $y$ defined as in \eqr{ydef}.

Now we turn our attention to the related problem of the nucleon spin
``crisis''. The nucleon valence-quark scheme \cite{BLP64} predicts
axial-vector spin components of the nucleon as
\begin{equation}
\Delta u_v \: = \: \frac{4}{3} \;\;\; , \;\;\;
\Delta d_v \: = \: -\frac{1}{3} \;\;\; , \;\;\;
\Delta s_v \: = \: 0 \; ,
\label{valence}
\end{equation}
where the subscript ``$v$'' refers to valence.
The EMC data of SLAC-Yale \cite{EMC88} improved this valence spin prediction
to \cite{AS89}
\begin{equation}
\Delta u \: = \: 0.94\pm0.007 \;\;\; , \;\;\;
\Delta d \: = \: -0.31\pm0.07 \;\;\; , \;\;\;
\Delta s \: = \: -0.02\pm0.07 \; .
\label{spinemc}
\end{equation}
Given the recent \cite{PDG06} $g_A=1.2695$ value deduced from neutron $\beta$
decay, and the axial baryon $d/f$ ratio $(d/f)_A\approx 1.74$ \cite{DF89},
with $d+f\equiv1$, one gets \cite{SKR06,BS03},
\begin{equation}
\Delta u \: = \: 0.870 \;\;\; , \;\;\;
\Delta d \: = \: -0.400 \;\;\; , \;\;\;
\Delta s \: = \: -0.057 \; .
\label{spindf}
\end{equation}
Dynamical tadpole leakage, using the axial-vector mesons $f_1(1285)$,
$f_1(1420)$ \cite{PDG06} and a reasonable mixing angle $\phi_A\sim15^\circ$,
leads to very similar predictions \cite{S84,BS03,SKR06}. These values are also
totally compatible with the QCD Bj\"{o}rken-sum-rule result (including
higher-twist effects) \cite{EK94}
\begin{equation}
\Delta u \: = \: 0.85\pm0.03 \;\;\; , \;\;\;
\Delta d \: = \: -0.41\pm0.03 \;\;\; , \;\;\;
\Delta s \: = \: -0.08\pm0.03 \; .
\label{spinqcd}
\end{equation}

We should stress the self-consistency between $\Delta m_N^{(s)}$ leading to
the $y$ values in Eqs.~(\ref{ysmall}--\ref{ytadpole}) --- including data in
\eqr{yexp} --- and the $|\Delta s|\sim6$\% scale in \eqrs{spindf}{spinqcd}.

The issue of the strange-quark content of the proton has received a great deal
of attention from chiral perturbation theory (ChPT) over the decades. First,
25 years ago, Gasser estimated \cite{G81} (also see Ref.~\cite{DN86}) a very
large value of
\begin{equation}
y \; \sim \; 60\,\mbox{\%} \; ,
\label{ygasser}
\end{equation}
still using the definition of \eqr{ydef}. Later, Gasser, Leutwyler and Sainio
\cite{GLS91} reduced the ChPT prediction to
\begin{equation}
y \; \sim \; 20\,\mbox{\%} \; ,
\label{ychpt}
\end{equation}
leading them to conclude: \em ``This value appears to be quite reasonable as
the corresponding contribution of the term $m_s\bar{s}s$ to the proton mass is
of order $(m_s/2\hat{m})\times10$ MeV $\approx130$ MeV.'' \em 
Still very recently \cite{S05}, Sainio held on to this ChPT value:
\em ``This value of $y$ corresponds to about 130 MeV in the proton mass being
due to the strange sea.'' \em In view of the above-mentioned new
experimental and lattice results, the present-day ChPT prediction appears to
be an order of magnitude too large, not to speak of the original ChPT
estimates. In contrast, the various other theoretical approaches outlined in
the foregoing are in agreement with the new data, not only for the strangeness
contribution to the proton mass, but also for its spin. Therefore, a
reassessment of some of the premises of ChPT seems inevitable.\footnote
{In \em Generalized \em \/ChPT (GChPT), the ratio $(m_s/\hat{m})_{\cur}$ could
be as small as $\approx\!10$, with $\hat{m}$ possibly as large as $\sim\!20$
MeV \cite{KS94}. This is to be contrasted with the standard ChPT values
$(m_s/\hat{m})_{\cur}\approx25\,$--$\,30$ MeV and $\hat{m}\sim5$ MeV. Thus,
GChPT would probably lead to a less serious discrepancy with the observed small
strangeness content of the proton.}

Finally, let us come back to the strongly related issue of the $\pi N$ 
$\sigma$-term in \eqr{dmns}, which has been at the centre of intense debate
for the past thirty odd years. To be more precise, on the experimental side
there has never been a real controversy about $\sigma_{\pi N}$, with measured
and extracted values clearly converging over the years: $58\pm13$ MeV in 1971
\cite{HJS71}, $66\pm9$ MeV in 1974 \cite{NO74}, $64\pm8$ MeV in 1982
\cite{K82}, and no significant deviations from the latter value in more recent
determinations. However, ChPT predictions of $\sigma_{\pi N}$ have been 
more volatile. In 1982, the review by Gasser and Leutwyler estimated it at
$24\,$--$\,35$ MeV \cite{GL82}. By 1991, ChPT had developed a much fancier
scheme \cite{GLS91}, in which three additional contributions were added to
the original, perturbative term due to Gell-Mann, Oakes and Renner (GMOR)
\cite{GMOR68}, and $\sigma_{\pi N}$ also ceased to be a $c$-number:
\begin{equation}
\sigma_{\pi N}(t=2m^2_\pi) \; = \; \sigma_{\pi N}^{\mbox{\scriptsize GMOR}} +
\sigma_{\pi N}^{\mbox{\scriptsize HOChPT}} +
\sigma_{\pi N}^{\bar{s}s} + \sigma_{\pi N}^{t-\mbox{\scriptsize dependent}}
\; \approx\; (25 + 10 + 10 + 15)\;\mbox{MeV} \; = \; 60 \; \mbox{MeV} \; .
\label{sigmachpt}
\end{equation}
Here, the second term on the right-hand side arises from higher-order ChPT,
the third one from the strange-quark sea, and the fourth is a $t$-dependent
contribution due to going from $t=0$ to the Cheng-Dashen \cite{CD70} point
$t=2m^2_\pi$, where the $\pi N$ background is minimal. Leutwyler \cite{L92}
concluded: \em ``The three pieces happen
to have the same sign.'' \em Of course, for things to work out right, all
\em four \em \/pieces must have the same sign, including the GMOR term.
However, by 1993 it had already become very clear from experiment \cite{R93}
that the strange-sea contribution to $\sigma_{\pi N}$ is not of the order of
10 MeV, but rather $\leq2$ MeV. The very recent combined experimental
\cite{T06a} and lattice \cite{T06b} results have now strengthened this body
of evidence. In complete accord with the latter picture and the experimental
value of $\sigma_{\pi N}$ are nonperturbative estimates based on the
infinite-momentum frame \cite{CSS91,SKR06}, predicting only \em one \em \/term,
viz.\
\begin{equation}
\sigma_{\pi N} \; \sim \; \frac{m^2_{\Xi} + m^2_\Sigma - 2 m^2_N}{2 m_N}
\left( \frac{m^2_\pi}{m^2_K-m^2_\pi} \right) \; \approx  \; 63 \; \mbox{MeV}\;,
\label{sigmaimf}
\end{equation}
or via sea contributions from a \em nonstrange \em \/$\sigma$ resonance
\cite{CSS93,SKR06}, yielding
\begin{equation}
\sigma_{\pi N} \; = \; \sigma_{\pi N}^{\mbox{\scriptsize quenched}} +
\sigma_{\pi N}^{\mbox{\scriptsize unquenched}} \; = \;
\sigma_{\pi N}^{\mbox{\scriptsize quenched}} + \frac{m^2_\pi}{m^2_\sigma}\,m_N
 \; \approx \; (25+40) \; \mbox{MeV} \; = \; 65 \; \mbox{MeV} \; ,
\label{sigmaunquenched}
\end{equation}
for $m_\sigma\approx650$ MeV, where the quenched contribution was determined
by quenched-lattice calculations \cite{APE90}.

\section*{Acknowledgments}
This work was supported by the {\it Funda\c{c}\~{a}o para a Ci\^{e}ncia
e a Tecnologia} \/of the {\it Minist\'{e}rio da Ci\^{e}ncia, Tecnologia
e Ensino Superior} \/of Portugal, under contract POCI/FP/63437/2005. 
One of us (F.K.) also acknowledges financial support from the Czech project
LC06002.

\end {document}